\documentclass[reprint,superscriptaddress,amsmath,amssymb,aps,prl]{revtex4-2}
\usepackage[colorlinks, linkcolor = ForestGreen, anchorcolor = blue, citecolor = magenta]{hyperref}
\usepackage{amsmath,amssymb,amsthm,mathtools,mathrsfs}
\usepackage[table,dvipsnames]{xcolor}
\usepackage{tikz}
\usepackage{adjustbox}
\usepackage{dsfont}
\usepackage[normalem]{ulem}
\usepackage{enumerate}
\usepackage{physics}

\theoremstyle{plain}
\newtheorem{theorem}{Theorem}
\newtheorem{lemma}{Lemma}
\newtheorem{proposition}{Proposition}

\theoremstyle{definition}
\newtheorem{definition}{Definition}

\newtheorem{example}{Example}

\newcommand{\bd}{\begin{definition}}
\newcommand{\ed}{\end{definition}}
\newcommand{\bt}{\begin{theorem}}
\newcommand{\et}{\end{theorem}}
\newcommand{\bn}{\begin{proposition}}
\newcommand{\en}{\end{proposition}}
\newcommand{\be}{\begin{equation}}
\newcommand{\ee}{\end{equation}}
\newcommand{\blem}{\begin{lemma}}
\newcommand{\elem}{\end{lemma}}
\newcommand{\bx}{\begin{example}}
\newcommand{\ex}{\end{example}}
\newcommand{\bprf}{\begin{proof}}
\newcommand{\eprf}{\end{proof}}

\DeclareMathAlphabet{\mathpzc}{OT1}{pzc}{m}{it} 
 \DeclareFontFamily{OT1}{pzc}{}
 \DeclareFontShape{OT1}{pzc}{m}{it}{ <-> s*[1.2] pzcmi7t }{}
 \DeclareMathAlphabet{\mathpzc}{OT1}{pzc}{m}{it}
 
\newcommand{\Ad}{\mathrm{Ad}}

\newcommand{\id}{\operatorname{id}}

\def\A{\mathfrak{A}}

\def\E{\mathcal{E}}
\def\H{\mathcal{H}}

\begin{document}									
\preprint{APS/123-QED}

\title{Assignment maps for arbitrary reduced quantum dynamics}

\author{James Fullwood}
\email{fullwood@hainanu.edu.cn}
\affiliation{School of Mathematics and Statistics, Hainan University, Haikou, Hainan Province, 570228, China}
\affiliation{Hainan International Exchange Center for Theoretical Physics, Haikou, Hainan Province, 570228, China}
\author{Weixiang Ye}
\email{wxy@hainanu.edu.cn}
\affiliation{Hainan International Exchange Center for Theoretical Physics, Haikou, Hainan Province, 570228, China}
\affiliation{Center for Theoretical Physics, School of Physics and Optoelectronic Engineering, Hainan University, Haikou, Hainan Province, 570228, China}

\date{\today}

\begin{abstract}
Given an open quantum system $S$ with environment $E$, we establish a one-to-one correspondence between completely positive trace-preserving (CPTP) maps from $S$ to $E$ and assignment maps initializing the composite system $SE$ in a possibly correlated state prior to unitary evolution. We show that every possible joint state $\rho_{SE}$ can be realized by such assignment maps, which are inherently Hermiticity-preserving and trace-preserving (HPTP). While these maps sacrifice global positivity, we derive a simple necessary and sufficient condition for any local state of $S$ to belong to their domain of validity. Furthermore, we prove that the geometry of this domain is governed by the joint state assigned to the maximally mixed input, ensuring that a non-zero volume of admissible local states can readily coexist with initial system-environment entanglement. Finally, we provide an operational protocol for witnessing such limits of validity.
\end{abstract}

	\maketitle
	
In the 17th century, Newton formulated the second law of motion, laying the foundation for the fundamental paradigm that the evolution of a physical system should be determined by dynamical laws subject to initial conditions. Such a paradigm manifests itself in the quantum domain via Schr\"{o}dinger's equation, which describes the unitary evolution of a closed quantum system under a Hamiltonian constraint. However, in practice quantum systems are rarely closed, as they are inevitably prone to interaction with their environment. 

Since the mid-1990s, physicists have been in disagreement about the appropriate mathematical description of the reduced dynamics of an open quantum system whose initial conditions include system-environment correlations~\cite{Pechukas_1994, Alicki_1995}. In the absence of such correlations, the composite system is initially in a product state $\rho_S\otimes \rho_E$,  where $S$ denotes the system of interest and $E$ denotes the environment. If we assume that the composite system $SE$ evolves unitarily under these initial conditions as governed by Schr\"{o}dinger's equation, then the reduced dynamics of system $S$ is described by the completely positive trace-preserving (CPTP) map corresponding to composition $\Tr_E\circ \text{Ad}_U\circ \A$, where $\Tr_E$ is the partial trace map which traces over the degrees of freedom of the environment $E$, $\Ad_U$ is the associated unitary evolution on the composite system $SE$, and $\A$ is the mapping which takes an arbitrary state $\rho_S$ to $\A(\rho_S)=\rho_S\otimes \rho_E$. Moreover, it follows from Stinespring's Theorem that every CPTP map is of this form~\cite{Stinespring_1955}.

For general reduced quantum dynamics, the essential ambiguity is what the analog of $\A$ should be when the composite system $SE$ is initially correlated. As proven by Pechukas and Jordan \emph{et al.}~\cite{Pechukas_1994, Jordan_2004}, every map from $S$ to $SE$ satisfying linearity, consistency ($\Tr_E\circ \A = \id_S$), and complete positivity must output an uncorrelated product state. Therefore, incorporating initial system-environment correlations requires relaxing at least one of these conditions. The point of view taken here is that relaxing positivity becomes natural upon distinguishing active physical transformations from passive transformations corresponding to a shift between local and global descriptions of a quantum system.  In particular, just as the partial trace $\Tr_E$ acts as a passive shift from the global description of $SE$ to the local description of $S$, an assignment map $\A$ should act as a `reverse shift' from the local description of $S$ back to the global description of $SE$. Upon adopting such a viewpoint, there is no reason to mandate that an assignment map $\A$ must correspond to a CPTP map. In other words, an assignment map is inherently kinematical rather than dynamical.

In accordance with this viewpoint, researchers who chose to retain the linearity and consistency of assignment maps recognized that the resulting reduced dynamics are inevitably described by maps that are not completely positive, but rather Hermiticity-preserving and trace-preserving (HPTP)~\cite{Shaji_2005, Carteret_2008, RodriguezRosario_2010, Modi_2012, Schmid_2019}. While HPTP dynamical maps have been analyzed across various frameworks~\cite{Buscemi_2014, Dominy_2016, VacchiniAmato_2016, Colla_2022, Settimo_2025}, their domains of validity have typically been determined on a case-by-case basis. As a result, a general, systematic construction of assignment maps for arbitrary open system dynamics remains absent from the literature.

In this Letter, we resolve this gap by constructing HPTP linear assignment maps capable of realizing every possible initial state $\rho_{SE}$. Remarkably, despite sacrificing global positivity, we establish a strict one-to-one correspondence between these kinematical maps and physical, CPTP channels. In particular, composing the associated channel with local transposition yields a kinematical mapping between the initial states of $S$ and $E$. Under this correspondence, each assignment map is uniquely characterized by a bipartite reference state corresponding to the joint state assigned to the maximally mixed input. We show that the local commutant of the projection onto this reference state's support determines the affine hull of the domain of validity, while positive semi-definiteness on this support bounds the allowed magnitude of local state perturbations.

Furthermore, this precise geometric description provides an explicit operational framework for prepare-evolve-measure scenarios involving open quantum systems in the presence of initial system-environment correlations. In particular, the boundary of the domain of validity serves as the mathematical manifestation of local preparation restrictions imposed by global correlations, defining the physically accessible states one can prepare in the lab. As a concrete illustration, we consider a family of assignment maps associated with qubit depolarizing channels parameterized by the depolarization strength $p$. For these maps, the admissible Bloch radius of the initial state of the open system increases as $\sqrt{p}$, demonstrating that a non-zero, three-dimensional volume of local preparations can readily coexist with an entangled reference state. Operationally, this non-zero volume of admissible states guarantees that the assignment map is robust to experimental noise, ensuring that small preparation errors will not inadvertently push the open system outside the bounds of physical compatibility.

\emph{Assignment maps via virtual quantum broadcasting.---} Let $S$ and $E$ denote quantum systems with finite-dimensional Hilbert spaces $\H_S$ and $\H_E$, and let $d_S=\dim(\H_S)$. The algebra of linear operators on a Hilbert space $\H$ will be denoted by $\mathcal{L}(\H)$, and the state space of $S$ by $\mathfrak{S}(S)=\{\rho_S\ge 0 : \Tr\rho_S = 1\}$. Given an ancillary copy $S'$ of $S$ identified through a fixed orthonormal basis $\{\ket{i}\}_{i=1}^{d_S}$, we let ${\tt SWAP}:\H_S\otimes \H_{S'}\to \H_S\otimes \H_{S'}$ denote the swap operator defined by ${\tt SWAP}\ket{i}\ket{j}=\ket{j}\ket{i}$. The mapping $\mathcal{B}:\mathcal{L}(\H_S)\to \mathcal{L}(\H_S)\otimes \mathcal{L}(\H_{S'})$ given by
\be
\mathcal{B}(\rho)=\frac{1}{2}\big\{\rho\otimes I_{S'},{\tt SWAP}\big\}\, ,
\ee
where $\{\cdot\, , \cdot\}$ denotes the anticommutator, will be referred to as the \emph{canonical virtual broadcasting map}. In Ref.~\cite{PFBC23}, it was shown that the canonical virtual broadcasting map is uniquely characterized by the following properties: (i) $\mathcal{B}(U\rho U^{\dag})=(U\otimes U)\mathcal{B}(\rho)(U\otimes U)^{\dag}$ for all unitaries $U$ on $\H_S$ and for all $\rho\in \mathcal{L}(\H_S)$. (ii) ${\tt SWAP}\mathcal{B}(\rho){\tt SWAP}=\mathcal{B}(\rho)$ for all $\rho\in \mathcal{L}(\H_S)$. (iii) If $\mathcal{D}:\mathcal{L}(\H_S)\to \mathcal{L}(\H_S)$ is completely dephasing with respect to the basis $\{\ket{i}\}$, then $\mathcal{B}_{\text{cl}}=(\mathcal{D}\otimes \mathcal{D})\circ \mathcal{B}\circ \mathcal{D}$ is the classical broadcasting map given by $\mathcal{B}_{\text{cl}}(\dyad{i}{j})=\delta_{ij}\dyad{i}{i}\otimes \dyad{i}{i}$. Properties (i), (ii) and (iii) are referred to as unitary covariance (UC), permutation invariance (PI), and classical consistency (CC), respectively. Furthermore, the map $\mathcal{B}$ satisfies the broadcasting condition $\Tr_1\circ\, \mathcal{B}=\Tr_2\circ\, \mathcal{B}=\id_S$, where $\Tr_k$ denotes the partial trace over the $k$th copy of $S$. The terminology `virtual' indicates that $\mathcal{B}$ is not a positive map. However, the canonical virtual broadcasting map is Hermiticity-preserving and trace-preserving (HPTP), and operational interpretations of it can be found in Refs.~\cite{PFBC23,FuPa24}.

Let $T_{S'}$ and $T_E$ denote transposition with respect to fixed orthonormal bases of $S'$ and $E$, and let $\mu=I_S/d_S$ denote the maximally mixed state of $S$. Then 
\be \label{MAXENTGLXD57}
(\id_S\otimes T_{S'})(\mathcal{B}(\mu))=\frac{1}{d_S}\sum_{i,j}\dyad{i}{j}\otimes \dyad{i}{j} = P^+\, ,
\ee
which is the projector onto the maximally entangled state $\ket{\phi^+}=d_S^{-1/2}\sum_i \ket{i}\otimes\ket{i}$. If we assume that $S$ and $E$ are equidimensional, and if we associate the copy $S'$ with $E$, then the mapping $\mathfrak{A}=(\id_S\otimes T_E)\circ \mathcal{B}$ may be viewed as an assignment map for the reduced dynamics of any unitary evolution of the composite system $SE$ initialized in the maximally entangled state $P^+$. In particular, the map $\A$ is linear, consistent in the sense that $\Tr_E\circ\, \A = \id_S$, and HPTP. Recall that the non-positivity of $\A$ does not concern us, as $\A$ is meant to be kinematical rather than dynamical.

In such a case, however, the map $\A$ is only positive when restricted to the singleton $\{\mu\}\subset \mathfrak{S}(S)$. At the other extreme, namely, when $SE$ is initially uncorrelated, we can construct the associated assignment map in a similar fashion. In particular, if we set $\mathfrak{A}=(\id_S\otimes (T_E\circ \E))\circ \mathcal{B}$, where $\E:\mathcal{L}(\H_{S'})\to\mathcal{L}(\H_E)$ is a discard-and-prepare channel of the form $\E(X)=\Tr(X)\rho_E^{T_E}$ for a fixed environment state $\rho_E$, then $\A(\rho_S)=\rho_S\otimes \rho_E$ for all $\rho_S\in \mathfrak{S}(S)$, thus the domain of validity of $\A$ is the entire state space $\mathfrak{S}(S)$.

Remarkably, every state $\rho_{SE}$ may be realized by an assignment map of the form $\mathfrak{A}=(\id_S\otimes (T_E\circ \E))\circ \mathcal{B}$ for some completely positive trace-preserving (CPTP) map $\E:\mathcal{L}(\H_{S'})\to \mathcal{L}(\H_E)$. Indeed, let $\rho_{SE}$ be an arbitrary state of $SE$, so that $\rho_{SE}$ is a density operator on the tensor product $\H_S\otimes \H_E$. By the Partial Transpose Theorem from Ref.~\cite{FuLi_2026}, there exists a CPTP map $\E:\mathcal{L}(\H_{S'})\to \mathcal{L}(\H_E)$ such that $\rho_{SE}^{T_E}=(\id_S\otimes\, \E)(\mathcal{B}(\rho_S))$, where $T_E$ denotes the partial transpose over system $E$ and $\rho_S=\Tr_E(\rho_{SE})$. Applying transposition to subsystem $E$ once more yields $\rho_{SE}=(\id_S\otimes (T_E\circ \E))(\mathcal{B}(\rho_S))$, demonstrating that $\A=(\id_S\otimes (T_E\circ \E))\circ \mathcal{B}$ realizes the initial conditions $\rho_{SE}$. Moreover, as $\rho_{E}=\Tr_S(\A_{\E}(\rho_S))=(T_E\circ \E)(\rho_S)$, it follows that $T_E\circ \E$ yields a kinematical mapping from initial states of $S$ to initial states of $E$.

\emph{A one-to-one correspondence.---} It follows from Eq.~\eqref{MAXENTGLXD57} that given \emph{any} CPTP map $\E:\mathcal{L}(\H_{S'})\to \mathcal{L}(\H_E)$, the prescription $\A_{\E}=(\id_S\otimes (T_E\circ \E))\circ \mathcal{B}$ yields an assignment map whose domain of validity includes the maximally mixed state $\mu$ of $S$. Moreover, we now show that the mapping $\mathcal{E}\mapsto \A_{\E}$ establishes a strict one-to-one correspondence between CPTP maps and this canonical family of assignment maps for reduced quantum dynamics. 

First, we show that distinct CPTP maps yield distinct assignment maps. Since the transpose operation is an involution (i.e., $T\circ T=\id$), applying $\id_S\otimes T_{S'}$ to both sides of Eq.~\eqref{MAXENTGLXD57} yields $\mathcal{B}(\mu)=(\id_S\otimes T_{S'})(P^+)$. Substituting this expression into the definition of the assignment map associated with a CPTP map $\E$ then yields
\begin{align*}
\A_{\E}(\mu)&= (\id_S\otimes (T_E\circ \E))\big( (\id_S\otimes T_{S'})(P^+) \big) \\
& = \big( \id_S\otimes (T_E\circ \E\circ T_{S'}) \big)(P^+) \\
&=\frac{1}{d_S}\mathscr{C}[T_E\circ \E\circ T_{S'}]\, ,
\end{align*}
where $\mathscr{C}[\Phi]=\sum_{i,j}\dyad{i}{j}\otimes \Phi(\dyad{i}{j})$ denotes the Choi matrix of a linear map $\Phi$~\cite{Ch75}. Moreover, since pre- and post-composing a CPTP map with transposition yields another CPTP map, Choi's Theorem guarantees that this normalized Choi matrix is positive semi-definite with unit trace. Consequently, $\A_{\E}(\mu)$ is a valid quantum state for every CPTP map $\E$. Furthermore, since two linear maps coincide if and only if their Choi matrices are equal, and since transposition is an involution, it follows by contraposition that if $\E\neq \E'$, then $\A_{\E}(\mu)\neq \A_{\E'}(\mu)$. Therefore, each CPTP map $\E$ defines a unique reference state $\A_{\E}(\mu)$, which implies that the mapping $\E\mapsto \A_{\E}$ is one-to-one.

Furthermore, by Choi's theorem, any joint density operator $R$ with local marginal $\mu$ is proportional to the Choi matrix of a unique CPTP map $\Phi_R$, which in turn defines a valid CPTP channel $\mathcal{E} = T_E \circ \Phi_R \circ T_{S'}$ satisfying $\mathfrak{A}_{\mathcal{E}}(\mu) = R$. It then follows that CPTP maps, bipartite reference states with marginal $\mu$, and this canonical family of assignment maps are all in strict one-to-one correspondence.

\emph{Determining the domain of validity.---} Given a CPTP map $\E:\mathcal{L}(\H_{S'})\to \mathcal{L}(\H_E)$, let $\text{dov}(\A_{\E})$ denote the domain of validity of the associated assignment map $\A_{\E}$, which consists of all states $\rho_S\in \mathfrak{S}(S)$ such that $\A_{\E}(\rho_S)$ is positive semi-definite. To streamline notation, let $R_{\E} = \A_{\E}(\mu)$ denote the joint reference state. It turns out that $\rho_S\in \text{dov}(\A_{\E})$ if and only if $\{ \rho_S\otimes \mathds{1}_E, R_{\E}\}$ is positive semi-definite, which provides a computable necessary and sufficient condition for belonging to $\text{dov}(\A_{\E})$. 

To derive this condition, we first observe that since $\A_{\E}$ is constructed via virtual broadcasting, its action on an arbitrary state $\rho_S$ may be expressed in terms of its action on the maximally mixed state $\mu$. Specifically, expanding the canonical virtual broadcasting map yields
\[
\A_{\E}(\rho_S) = (\id_S\otimes (T_E\circ \E))\left( \frac{1}{2}\{\rho_S\otimes \mathds{1}_{S'}, {\tt SWAP}\} \right) \, .
\]
Since the operator $\rho_S\otimes \mathds{1}_{S'}$ acts trivially on the second subsystem, left- and right-multiplication by this operator commutes with the local map $\id_S\otimes (T_E\circ \E)$. Pulling $\rho_S$ outside the argument of the map yields
\begin{align*}
\A_{\E}(\rho_S)&= \frac{1}{2}\big\{ \rho_S\otimes \mathds{1}_E, (\id_S\otimes (T_E\circ \E))({\tt SWAP}) \big\} \\
&= \frac{d_S}{2}\big\{ \rho_S\otimes \mathds{1}_E, R_{\E} \big\} \, ,
\end{align*}
where the final equality follows from the fact that ${\tt SWAP} = d_S \mathcal{B}(\mu)$. As such, it follows that $\rho_S\in \text{dov}(\A_{\E})$ if and only if $\{ \rho_S\otimes \mathds{1}_E, R_{\E}\} \geq 0$. Moreover, since two commuting positive operators always have a positive product, it follows that $[\rho_S\otimes \mathds{1}_E,R_{\E}]=0$ is a sufficient condition for validity.

We now utilize this result to formulate a mathematically precise relationship between the geometry of $\text{dov}(\A_{\E})$ and the reference state $R_{\E}$. Let $\rho_S\in \text{dov}(\A_{\E})$, let $\Pi_{\E}$ denote the orthogonal projector onto the support of $R_{\E}$, and consider the direct sum decomposition $\H_S\otimes \H_E = \text{supp}(R_{\E}) \oplus \text{ker}(R_{\E})$. With respect to this decomposition, there exists a positive definite operator $P$ and operators $X$ and $Y$ such that
\[
R_{\E} = \begin{pmatrix} 
P & 0 \\ 
0 & 0 
\end{pmatrix} \quad \text{and} \quad \rho_S\otimes \mathds{1}_E = 
\begin{pmatrix} 
X & Y \\ 
Y^\dag & Z 
\end{pmatrix}\, ,
\] 
which implies
\[
\big\{ \rho_S\otimes \mathds{1}_E, R_{\E} \big\} = \begin{pmatrix} XP + PX & PY \\ Y^\dag P & 0 \end{pmatrix} \, .
\]
Since $\rho_{S}\in \text{dov}(\A_{\E})$, this anticommutator must be positive semi-definite. Evaluating the quadratic form of this anticommutator on a vector $u \oplus tv$ (where $u \in \text{supp}(R_{\E})$, $v \in \text{ker}(R_{\E})$, and $t$ is an arbitrary complex number) contains $2\operatorname{Re}(t\,u^\dag PYv)$ but no term quadratic in $t$. Positivity therefore forces the off-diagonal blocks $PY$ and $Y^{\dag}P$ to vanish. Since $P$ is positive definite it is invertible, which forces $Y = 0$. Therefore, $\rho_S\otimes \mathds{1}_E$ must be block-diagonal with respect to the support of $R_{\E}$, which is equivalent to the commutation relation 
\be\label{eq:commutant_bound}
[\rho_S\otimes \mathds{1}_E, \Pi_{\E}] = 0\, .
\ee
While Eq.~\eqref{eq:commutant_bound} determines the affine hull of $\text{dov}(\A_{\E})$, the actual boundaries of the domain are determined by the positivity of the anticommutator restricted to the support of $R_{\E}$, which further requires $XP + PX \geq 0$.

This commutant constraint uniquely fixes the affine hull of the domain of validity. In particular, if we define the real vector space $\mathcal{V}_{\E}$ consisting of all traceless Hermitian operators $H$ on $\H_S$ satisfying $[H \otimes \mathds{1}_E, \Pi_{\E}] = 0$, then the affine hull of $\text{dov}(\A_{\E})$ is precisely $\mu + \mathcal{V}_{\E}$. Equation~\eqref{eq:commutant_bound} then dictates that all valid states must lie within this affine subspace. Conversely, to see that the domain possesses a non-empty relative interior within this subspace---rather than degenerating into a lower-dimensional subset---let $\lambda_+(R_{\E})$ be the smallest strictly positive eigenvalue of $R_{\E}$, and let $\|\cdot\|_\infty$ denote the operator norm. For any $H \in \mathcal{V}_{\E}$ satisfying the bound
\be
\|H\|_\infty \leq \frac{\lambda_+(R_{\E})}{d_S\|R_{\E}\|_\infty}\, ,
\ee
the state $\mu + H$ is positive semi-definite, and commutation with $\Pi_{\E}$ confines the perturbation $(d_S/2)\{H \otimes \mathds{1}_E, R_{\E}\}$ entirely to the support, preserving the positivity of the assigned joint state. Therefore, $\text{dov}(\A_{\E})$ always contains a neighborhood of $\mu$ inside $\mu + \mathcal{V}_{\E}$ provided the subspace $\mathcal{V}_{\E}$ is non-trivial.

The dimension of $\mathcal{V}_{\E}$ counts the independent directions of the system state compatible with the assignment, fundamentally tying local preparation freedom to global correlation. For instance, if the reference state $R_{\E}$ is pure, its local marginal $\mu$ dictates that it must be maximally entangled. The commutant condition~\eqref{eq:commutant_bound} then forces the pure state vector to be an eigenvector of $\rho_S \otimes \mathds{1}_E$, and full Schmidt rank forces $\rho_S$ to be proportional to $\mathds{1}_S$, collapsing the domain to the single point $\{\mu\}$. Conversely, if $R_{\E} = \mu \otimes \rho_E$ is an uncorrelated product state, $\Pi_{\E} = \mathds{1}_S \otimes \Pi_E$, which trivially commutes with any $\rho_S \otimes \mathds{1}_E$. In such a case the commutant condition~\eqref{eq:commutant_bound} imposes no restriction, allowing the domain to encompass the entire state space $\mathfrak{S}(S)$. Furthermore, for a qubit system ($d_S=2$), the Pauli commutation relations dictate a strict geometric trichotomy: if $H_1, H_2 \in \mathcal{V}_{\E}$ are independent, then $i[H_1, H_2]$ provides a third independent operator in $\mathcal{V}_{\E}$. Consequently, the affine dimension of a qubit domain can only be $0$, $1$, or $3$ (a point, a line segment, or a convex body with non-zero volume).

For a concrete illustration, suppose $S$ and $E$ each consist of a single qubit with identified computational bases, and consider the depolarizing channel 
\[
\E_p(X) = (1-p)X + p\Tr(X)\frac{\mathds{1}_E}{2}\, ,
\]
where $p \in [0, 1]$ is the depolarizing parameter. Since the depolarizing channel is covariant with respect to transposition (i.e., $T_E\circ \E_p\circ T_{S'} = \E_p$), its reference state is 
\[
R_p = \A_{\E_p}(\mu) = (1-p)P^+ + \frac{p}{4}(\mathds{1}_S\otimes \mathds{1}_E)\, .
\]
To determine $\text{dov}(\A_{\E_p})$, we consider the Bloch vector representation of an arbitrary state $\rho_S = \frac{1}{2}(\mathds{1}_S + \vec{r}\cdot \vec{\sigma})$, so that $r = |\vec{r}|\leq 1$. Invariance of $R_p$ under local unitaries of the form $V \otimes V^*$ permits the alignment of $\vec{r}$ with the $z$-axis without changing the assigned state's spectrum, which consists of the four eigenvalues 
\be
\frac{p(1\pm r)}{4} \quad \text{and} \quad \frac{2-p \pm \sqrt{(2-p)^2 r^2 + 4(1-p)^2}}{4} \, .
\ee
Both signs occur in each pair. The first pair is non-negative for all valid $r \leq 1$, meaning the domain is entirely fixed by the smaller eigenvalue of the second pair. Enforcing positivity reveals $\rho_S\in \text{dov}(\A_{\E_p})$ if and only if
\be \label{BLOCH79}
r \leq r_{\max}(p) = \frac{\sqrt{p(4-3p)}}{2-p}\, .
\ee

The above inequality provides a precise geometric picture of how $\text{dov}(\A_{\E_p})$ expands as noise degrades the reference state. At the noiseless extreme ($p=0$), $R_0$ is pure and maximally entangled, forcing $r_{\max}(0) = 0$ and collapsing the domain to $\{\mu\}$. At the fully depolarizing extreme ($p=1$), $R_1$ is an uncorrelated product state, yielding $r_{\max}(1) = 1$ and encompassing the entire Bloch sphere. 

Most remarkably, as $p \to 0^+$, the admissible radius expands as $r_{\max}(p) = \sqrt{p} + O(p^{3/2})$. Arbitrarily weak depolarization thus yields a domain of full affine dimension (a 3D ball). Meanwhile, the partial transpose of $R_p$ possesses a negative eigenvalue $(3p-2)/4$ on the antisymmetric subspace, meaning $R_p$ remains entangled via the Peres-Horodecki criterion for all $0 < p < 2/3$~\cite{Peres_1996}. A three-dimensional volume of compatible system inputs therefore readily coexists with entanglement in the reference state.

\emph{Witnessing the limits of physical compatibility.---} In the presence of initial system-environment correlations, the assignment map $\A_{\E}$ modeling the reduced dynamics is uniquely determined by the joint reference state $R_{\E}$. While $R_{\E}$ can be obtained via quantum state tomography or theoretically specified, its central operational consequence is that Alice is not free to prepare any arbitrary state of $S$, as her physically accessible preparations are strictly confined to the domain of validity $\text{dov}(\A_{\E})$. Unless the reference state is pure and maximally entangled, this domain possesses a non-zero volume, guaranteeing that the assignment map is robust to experimental noise.

To operationally witness the limits of this physical compatibility, consider a linear trajectory of local preparations $\rho_S(t) = (1-t)\mu + t\rho_*$ with $t\in [0,1]$, where $\mu$ is the maximally mixed state and $\rho_*$ lies on the boundary of $\text{dov}(\A_{\E})$. At the boundary point $\rho_*$, the assigned joint state $\A_{\E}(\rho_*)$ attains a zero eigenvalue with corresponding normalized eigenvector $\ket{\psi_*}$. Performing the joint binary projective measurement $\{Q,\mathds{1}_{SE}-Q\}$ with $Q=U\dyad{\psi_*}{\psi_*}U^\dagger$ after a joint unitary evolution $U$ yields the theoretical Born rule probability
\begin{align*}
p_Q(t)&=\Pr[\text{outcome }Q\mid\rho_S(t)]=\Tr\!\left[Q\,U\A_{\E}(\rho_S(t))U^\dagger\right]\\
&=\bra{\psi_*}\A_{\E}(\rho_S(t))\ket{\psi_*}\\
&=(1-t)\bra{\psi_*}\A_{\E}(\mu)\ket{\psi_*} + t\bra{\psi_*}\A_{\E}(\rho_*)\ket{\psi_*}\, .
\end{align*}

Since $\ket{\psi_*}$ resides in the kernel of the boundary state $\A_{\E}(\rho_*)$, we have $\bra{\psi_*}\A_{\E}(\rho_*)\ket{\psi_*}=0$, from which it follows that $p_Q(t) = (1-t)p_Q(0)>0$ for all $0\leq t<1$. In particular, $p_Q(t)$ decreases monotonically with $t$ until finally reaching $0$ at $t=1$ when the joint state reaches the boundary of $\text{dov}(\A_{\E})$. Therefore, along this linear trajectory, $p_Q(t)$ directly quantifies the separation between the preparation $\rho_{S}(t)$ and the boundary of physical compatibility. In particular, small but non-zero values of $p_Q(t)$ indicate that the local preparation is on the verge of breaking the initial system-environment correlations modeled by $\A_{\E}$.

\emph{Concluding remarks.---} Despite decades of work on the mathematical description of reduced quantum dynamics, a systematic construction of assignment maps capable of realizing every correlated state of the system-environment remained absent from the literature. The framework presented here rigorously resolves this gap, providing an explicit construction of linear assignment maps via virtual quantum broadcasting. By showing that a quantum channel composed with local transposition yields a kinematical rather than a dynamical mapping between the initial states of $S$ and $E$, our construction establishes a one-to-one correspondence between assignment maps and physical CPTP maps, thus elucidating the physical origins of non-positive reduced dynamics. Moreover, our construction yields a precise geometric description of the domain of validity of a given assignment map, and we provided an operational protocol for witnessing the limits of validity. As such, our results not only resolve long-standing limitations in the theory of assignment maps, but also provide a concrete mathematical tool for analyzing general open-system dynamics where initial system-environment entanglement cannot be ignored.

\emph{Acknowledgments.---} J.F. is supported by the National Natural Science Foundation of China (W2632013), and by the Hainan Provincial Natural Science Foundation of China (126MS0010). J.F would also like to thank Seok Hyung Lie for fruitful discussion on the subject. W.Y. is supported by National Natural Science Foundation of China (62475062), and the Humboldt Research Fellowship Programme for Experienced Researchers (CHN-1218456-HFST-E).

\emph{AI usage declaration.---} During the preparation of this manuscript, the authors utilized Google Gemini for language refinement, and as an interactive tool for the exploration of analytical methods and mathematical derivations. The authors rigorously reviewed and edited all generated text and assume full responsibility for the content and technical accuracy of this work.

\bibliography{references}

\end{document}